\documentclass[seceq,preprint]{ptptex}
\usepackage{graphicx,bm,color,amsmath} 
\usepackage{wrapft}

\preprintnumber[3cm]{%<-- [..]: optional width of preprint # column.
OCU-PHYS-360\\AP-GR-95}

\title{Dynamical instability in a relativistic cylindrical shell composed of 
counter rotating particles}

\author{Yasunari \textsc{Kurita}$^{1,}$\footnote{E-mail: kurita@gen.kanagawa-it.ac.jp}
and Ken-ichi \textsc{Nakao}$^{2,}$\footnote{E-mail: knakao@sci.osaka-cu.ac.jp}
}

\inst{ 
$^1$Kanagawa Institute of Technology, Atsugi, Kanagawa 243-0292, Japan 
\\
$^2$Department of Mathematics and Physics, Graduate School of Science, 
   Osaka City  University, Osaka 558-8585, Japan 
} 

\abst{
We give a perturbative analysis for an infinitesimally thin cylindrical 
shell composed of counter rotating collisionless particles, originally devised 
by Apostolatos and Thorne.
They found a static solution of the shell and concluded by C-energy argument that it is stable.
Recently, the present authors and Ida reanalyzed this system by 
evaluating the C-energy on the future null infinity and found that 
the system has an instability, though it was not shown how the system 
is unstable. 
In this paper, it is shown in the framework of the linear perturbation theory 
that, if the constituent particles move slowly, 
the static shell is unstable in the sense that the perturbation of its 
circumferential radius oscillates with exponentially 
growing amplitude, whereas if the speed of the constituent particle exceeds a 
critical value, the shell just expands or contracts exponentially with time. 
}

\begin{document}
\maketitle

%==============================================================
\section{Introduction}
\label{sec:intro}
%============================================================== 

Nowadays, it may be a matter of time before one makes direct observations of 
gravitational waves, which will be a great success in experimental physics. 
Thus, it becomes more important to theoretically 
%On one hand, as a problem of theoretical physics, it is hard to say that we 
understand well about spacetime dynamics accompanied with 
emission of the gravitational radiation. However, it is not so easy task, since  
it is, in general, difficult to solve the Einstein equations 
because of their nonlinearity. 
One efficient way is to impose a symmetry on spacetime. 
The simplest one is the spherical symmetry, but there is no 
freedom  of gravitational radiation in spherically symmetric system, 
and hence this assumption is not adequate to study the effects of gravitational 
radiation. As another simple symmetry, cylindrical symmetry  
has been sometimes considered.
The cylindrically symmetric spacetime 
has a degree of freedom of gravitational radiation 
known as Einstein-Rosen gravitational waves\cite{ER-wave,Beck}.
The cylindrical system has been studied in connection with the 
gravitational waves by some authors\cite{Marder,Stachel,Piran}.

Apostolatos and Thorne 
studied an infinitesimally thin cylindrical shell composed of counter rotating 
collisionless particles with vanishing total angular momentum\cite{AT}. 
Hereafter, we refer to this as the Apostolatos-Thorne(AT)-shell. 
The AT-shell admits a static configuration, and 
Apostolatos and Thorne investigated its stability by the argument based on the  
C-energy which is a quasi-local energy per unit length 
in the translationally invariant direction\cite{Thorne:1965}.  
They concluded that the static configuration of the AT-shell 
is stable, and any dynamical configuration will finally settle down into 
the static state by releasing the energy through the gravitational emission.  
By contrast, the present authors and Ida\cite{Nakao:2007rb} have shown that 
some class of momentarily 
static and radiation free initial configurations does not settle down 
into static equilibrium configurations, or otherwise infinite amount 
of the C-energy through the gravitational radiation 
is released to the future null infinity.
It implies the existence of an instability in this system.
It should be noted that 
the C-energy argument by Apostolatos and Thorne 
is restricted in a bounded domain, whereas the argument in Ref.~[\ref{NIK}] 
is based on the conservation of the C-energy in the infinite domain. 

Recently, Gleiser and Ramirez\cite{Gleiser:2011yq} have
studied some aspect of the relativistic dynamics of the AT-shell. 
They have analytically solved the equation of motion for the shell 
and the Einstein equations by imposing a condition that the interior of 
the AT-shell is always flat. 
They have also investigated a perturbation around the static
configuration 
%with incoming gravitational radiation from past null infinity
and found that there are oscillating solutions for any chosen period.
It should be noted that they analyzed the cases which permit the   
C-energy input to the shell from the past null infinity and  
is different from the situation considered in Ref.~[\ref{NIK}].

In this paper, we reanalyze the AT-shell model in the framework 
of the linear perturbation theory.  We are interested in stability of the AT-shell 
in the case that there is no C-energy input to the shell from infinity, 
and hence we assume that there is no incoming gravitational 
radiation from past null infinity.

Hamity, C$\acute{\mbox{e}}$cere and  Barraco\cite{Hamity:2007jh} 
discussed the stability of the static state of the shell, 
and concluded that
there are both stable and unstable static configurations for the AT-shell. 
We will mention their result in connection with our present study 
in the last section.

This paper is organized as follows. 
In section \ref{section:construction-of-shell}, 
we describe the geometry of the cylindrical shell and construct 
the AT-shell model by using the coordinate system comoving to the AT-shell. 
In section \ref{section-static}, we describe the static solution for 
the shell.
In section \ref{section-perturbation},
we study the stability of the static AT shell by a perturbative analysis. 
Finally, a brief summary of the main results
is given in section \ref{section-summary}.

We adopt the geometrized units $c=G=1$.

%============================================================== 
\section{Cylindrical shell composed of counter-rotating particles} 
\label{section:construction-of-shell}
%============================================================== 

In this section, we give a description of the AT-shell model which is, as mentioned, 
an infinitesimally thin cylindrical shell composed of 
counter rotating collisionless particles\cite{AT}. 
For our present purpose, we adopt the coordinate system comoving to 
the shell, in which the spatial coordinates of the shell are kept constant.
This coordinate system is different from the 
original one given by Apostolatos and Thorne\cite{AT}.  

%%%%%%%%%%%%%%%%%%%%%%%%%%
\begin{figure}[htbp]
  \centering
  \includegraphics[width=0.3\linewidth,clip]{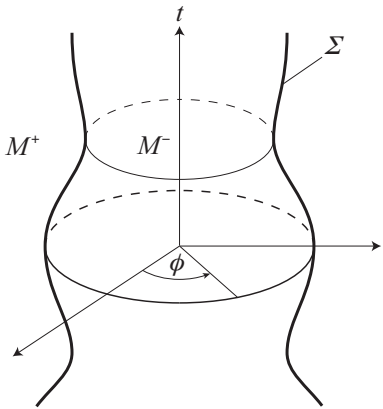}
  \caption{The world volume $\Sigma$ of the AT-shell divides the 
  spacetime into two regions $M_\pm$. }
  \label{fig:AT-shell}
\end{figure}
%%%%%%%%%%%%%%%%%%%%%%%%%%

The AT-shell divides the spacetime into two regions: one is the inside of 
the AT-shell $M^-$ and the other is the outside of it $M^+$ (see Fig.~\ref{fig:AT-shell}). 
By denoting the world volume of the AT-shell by $\Sigma$, the whole spacetime is 
given by $M=M^- \cup \Sigma\cup M^+$.  
Both regions $M^\pm$ have the whole 
cylinder symmetry, or in other words, the line element is written in the form, 
\begin{eqnarray}
ds^2 = e^{2(\gamma-\psi)} (-dt^2+dr^2) +\beta^2 e^{-2\psi}d\phi^2
+ e^{2\psi} dz^2,
\label{eq:metric-1}
\end{eqnarray}
where $\gamma$, $\psi$ and $\beta$ are functions 
that depend only on $t$ and $r$. 
We assume that both regions $M^\pm$ are vacuum. 
The domains of coordinates are $-\infty < t<\infty$, $-\infty<z <\infty$, $0\le r < \infty$,
and $0\le \phi < 2\pi$.
The axis of rotational symmetry is located at $r=0$.
The coordinate bases $\partial/\partial \phi$ and $\partial/\partial z$
are the rotational and translational Killing vectors, respectively.

The Einstein equations in the vacuum region lead to the equation for 
$\beta$, $\psi$ and $\gamma$ as
\begin{eqnarray}
&&  \beta''-\ddot{\beta} =0, \\
&&
\beta  (\ddot{\psi}-\psi'')+\dot{\beta}\dot{\psi}-\beta'\psi' = 0, \\
&& (\dot{\beta}^2-\beta'^2)\gamma'+\beta\beta' (\psi'^2+\dot{\psi}^2)
-2\beta\dot{\beta}\dot{\psi}\psi'+(\beta'\beta''-\dot{\beta}\dot{\beta}') 
=0 
 \\
&&(\beta'^2-\dot{\beta}^2)\dot{\gamma}+ \beta\dot{\beta}(\psi'^2+\dot{\psi}^2)
-2\beta\beta'\dot{\psi}\psi'  +(\dot{\beta}\beta''-\beta'\dot{\beta}') 
=0 \\
&& \ddot{\gamma}-\gamma''+(\dot{\psi}^2-\psi'^2)  =0,
\end{eqnarray}
where $\dot{A}$ and $A'$ mean derivative of $A$ 
with respect to $t$ and $r$, respectively.
In the original description of the AT shell, $\beta$ is set to be $r$, 
i.e., $\beta=r$. By contrast, we do not set it here 
so as to introduce the coordinate system comoving to  the shell later.

The stress-energy tensor of the AT-shell is infinite since the finite energy is confined 
within the infinitesimally thin region. The Ricci tensor diverges on the AT-shell 
through the Einstein equations, and hence the AT shell is the so-called 
s.p.~curvature singularity\cite{HE}. 
However, since the co-dimension of a shell is one, 
this singularity is so weak that the metric is defined on the AT-shell. 
Hence, even if the spacetime is singular on the AT-shell,
it can consistently be treated by the Israel's metric junction method 
\cite{Israel-1,Israel-2,Israel-3}. 
The junction conditions between $M^-$ and $M^+$ at $\Sigma$ 
require the continuity of the metric and specify the discontinuity of 
the extrinsic curvature $K^{\pm}_{\alpha\beta}$ compatible with the stress-energy 
tensor of the AT-shell. 

We require that the Killing vectors $\partial/\partial \phi$ and 
$\partial/\partial z$ are continuous at $\Sigma$, and this requirement leads to 
the continuities of the metric functions $\beta$ and $\psi$. 
We do not require the continuity of $\gamma$, and this means 
that the coordinate function $t$ and accordingly, the coordinate basis, 
$\partial/\partial t$, may not be continuous at the $\Sigma$. 
Although we can require that the coordinate 
function $t$ and accordingly, the coordinate basis, 
$\partial/\partial t$ is also continuous at $\Sigma$, we will not do so in this paper. 
This means that $\gamma$ may not be continuous at the AT-shell. 
The reason of this choice is that 
the solutions for the metric functions do not take simple forms 
in the continuous time coordinate, whereas   
the discontinuous one makes those solutions simple. 
The time coordinate for the interior region 
is denoted by $t_-$, whereas that for the exterior region is denoted by $t_+$. 
The radial coordinate at the AT-shell is denoted by $R(\tau)$, 
where $\tau$ is the proper time of an observer at rest on the shell.

We introduce the proper reference frame of an observer riding on the AT-shell
as follows:
\begin{eqnarray}
&&{\bf e}_{(\tau)} = \frac{d}{d\tau} =X_{\pm}\frac{\partial}{\partial 
 t_{\pm}}+V\frac{\partial}{\partial r}={\mbox{four velocity of the shell}}, \\
&&{\bf e}_{(n)}    =\frac{d}{dn} = V \frac{\partial}{\partial t_{\pm}}+ 
 X_{\pm} \frac{\partial}{\partial r}={\mbox{outward unit vector normal 
 to the shell}}, \\
&&{\bf e}_{(\phi)} = \frac{1}{\beta e^{-\psi}} \frac{\partial}{\partial \phi}, \\
&&{\bf e}_{(z)}    = \frac{1}{e^{\psi}} \frac{\partial}{\partial z},
\end{eqnarray}
where 
\begin{eqnarray}
V := \frac{dR}{d\tau}, \quad \mbox{and} \quad 
X_{\pm}:=\frac{d t_{\pm}}{d\tau} = \sqrt{e^{-2(\gamma^{\pm}-\psi)}+V^2}.
\label{eq:X}
\end{eqnarray}
The subscripts or superscripts $+$ and $-$ are used to 
denote quantities evaluated on the 
outer and inner faces of the shell, respectively.

Each constituent particle of the AT-shell has an identical rest mass.
Half of the particles orbit around the symmetry axis in a right-handed
direction with angular momentum per unit mass $\alpha$,
and the other half orbit in the opposite, left-handed direction with 
angular momentum per unit rest mass $-\alpha$.
Therefore, the net angular momentum of the AT-shell is zero.
%The circumferential radius ${\cal R}$ at the AT-shell and 
The absolute value of the specific  linear 
momentum $u$ of a constituent particle, which is defined by
\begin{eqnarray}
u:= \frac{\alpha}{\beta e^{-\psi}}, 
\end{eqnarray}
is an important quantity to describe the AT-shell. 
The other quantity characterizing the AT-shell is the shell's rest mass 
per unit translational Killing length, which is denoted by $\lambda$.
The rest mass of each constituent particle is conserved, 
so is $\lambda$.
Then, the surface stress-energy tensor of the AT-shell is given by
\begin{eqnarray}
{\bf S} = \sigma {\bf e}_{(\tau)}\otimes {\bf e}_{(\tau)} + T {\bf 
 e}_{(\phi)}\otimes {\bf e}_{(\phi)},
\end{eqnarray}
where we have defined 
the energy per unit area and the surface stress of the AT-shell as
\begin{eqnarray}
\sigma := \frac{\lambda\sqrt{1+u^2}}{2\pi \beta  }, 
\quad \mbox{and} \quad 
T:=\frac{\lambda u^2}{2\pi \beta  \sqrt{1+u^2}},
\end{eqnarray}
respectively.

Israel has shown that the Einstein equations 
on the AT-shell reduce to
\begin{eqnarray}
K^+_{\alpha\beta}-K^-_{\alpha\beta} = 
-8\pi\left(S_{\alpha\beta}-\frac{1}{2}S^{\mu}_{\mu} h_{\alpha\beta} \right),
\end{eqnarray}
where $K^{\pm}_{\alpha\beta}$ is the extrinsic curvature of the shell's 
outer and inner face, 
and $h_{\alpha\beta}$ is the induced metric on the $\Sigma$.

Now, we impose the comoving condition for the AT-shell, i.e., $V=0$. 
The comoving condition fixes the metric function $\beta$, 
but it does not necessarily imply $\beta=r$. 
In this coordinate system, the coordinate radius $R$ at the AT-shell does not 
characterize the evolution of the shell, 
but its circumferential radius $\cal R$ defined by
\begin{equation}
\mathcal{R}:= \beta e^{-\psi} |_{r=R},
\end{equation}
does instead of $R$. 

In the comoving coordinate, the non-vanishing components of 
the extrinsic curvature are given by
\begin{eqnarray}
&& K_{(\tau)(\tau)}=  \psi_{,n}-\gamma_{,n},  \\
&& K_{(\phi)(\phi)} = (\ln\beta)_{,n} -\psi_{,n}, \\
&& K_{(z)(z)} = \psi_{,n}.
\end{eqnarray}
Therefore, the Israel's junction conditions give the following equations:
\begin{eqnarray}
\left[e^{-\gamma}e^{\psi}\left(\psi'- \gamma' \right)\right]^{+}_{-}
&=& -\frac{2\lambda ( 1+2u^2)}{\beta  \sqrt{1+u^2}}, \\
\left[e^{-\gamma}e^{\psi}\left(\frac{\beta'}{\beta}-\psi'\right)\right]^{+}_{-} 
&=& -\frac{2\lambda (1+2u^2)}{\beta  \sqrt{1+u^2}}, \\
\left[e^{-\gamma}e^{\psi}  \psi'  \right]^{+}_{-} 
&=& -\frac{2\lambda}{\beta \sqrt{1+u^2}},
\end{eqnarray}
where 
\begin{equation}
\left[A\right]^+_-=A^+-A^-.
\end{equation}
Using the continuity of $\psi$ and $\beta$ at $\Sigma$, 
we can rewrite the above equations as
\begin{eqnarray}
\left[e^{-\gamma} \psi' \right]^{+}_{-}
&=& -\frac{2\lambda }{\beta  \sqrt{1+u^2}} e^{-\psi},  \label{eq:junction-1} \\
\left[e^{-\gamma}\gamma'  \right]^{+}_{-} 
&=& \frac{4\lambda u^2}{\beta  \sqrt{1+u^2}}e^{-\psi}, \label{eq:junction-2} \\
\left[e^{-\gamma}  \beta'  \right]^{+}_{-} 
&=& -4\lambda \sqrt{1+u^2} e^{-\psi}. \label{eq:junction-3} 
\end{eqnarray}
These are the junction conditions under the comoving condition.

%=================================================
\section{static solution}
\label{section-static}
%===================================================

As was found by Apostolatos and Thorne\cite{AT}, there is a static solution for 
the AT-shell which is regular except at the shell itself. 
In the interior region, the metric functions of the static solution are 
\begin{eqnarray}
\beta^-=r, \quad \psi^- = \psi_s, \quad \gamma^-=0,
\label{eq:zero-interior-solution}
\end{eqnarray}
where $\psi_s$ is a constant.
%Then, the interior metric has the following form:
%\begin{eqnarray}
%ds^2 = e^{-2\psi_s} (-dt^2_-+dr^2) +r^2 e^{-2\psi_s} d\phi^2+ e^{2\psi_s}dz^2.
%\end{eqnarray}
%Clearly, this solution is regular at the symmetry axis. 
In the exterior region, the metric functions are
\begin{eqnarray}
\beta^+ = r, \quad \psi^+ = \psi_s -2y \ln\frac{r}{R}, \quad 
\gamma^+ = 2\ln(1+2y) + 4y^2\ln\frac{r}{R},
\label{eq:0th-exterior-solution}
\end{eqnarray}
where we have defined a parameter 
\begin{equation}
y:=\frac{\alpha^2}{R^2e^{-2\psi_s}}.
\end{equation}
The junction conditions require the relation 
\begin{eqnarray}
\lambda e^{-\psi_s} = \frac{y\sqrt{1+y}}{(1+2y)^2}. 
\end{eqnarray}
Since we may set $\psi_s=0$ by rescaling the coordinates 
$t$, $r$ and $z$, the static solutions form  a one parameter family. 
If we fix $y$, then the solution is uniquely determined, and hence 
$y$ is a good parameter to characterize the static solutions. 

%=====================================================
\section{Perturbative analysis around the static solution}
\label{section-perturbation}
%======================================================

In this section, we give a perturbative analysis of the static AT-shell. 
If there is a smooth one-parameter family of dynamical exact solutions 
for the AT-shell,  which includes a static solution as its member, 
we can construct dynamical solutions from the static solution 
by the perturbative method\cite{Wald}.  
Here, we assume that such a one-parameter family exists. Denoting 
the parameter by $\varepsilon$, we write the metric functions in the form
\begin{eqnarray}
&& \beta(t,r;\varepsilon) = \sum_{n=0}\varepsilon^n\beta_n(t,r), \\
&& \psi(t,r;\varepsilon) = \sum_{n=0}\varepsilon^n\psi_n(t,r), \\
&& \gamma(t,r;\varepsilon) = \sum_{n=0}\varepsilon^n\gamma_n(t,r).
\end{eqnarray}
The static solution is obtained by putting $\varepsilon=0$, or in other words, 
$\beta_0$, $\psi_0$ and $\gamma_0$ are functions given by 
(\ref{eq:zero-interior-solution}) and (\ref{eq:0th-exterior-solution})
for the interior and exterior regions, respectively. 
We can derive the Einstein equations of each order with respect to $\varepsilon$, 
and solve them by successive approximation up to the order we need.
In this paper, we solve the equations up to $O(\varepsilon)$, i.e.,  
the linearized Einstein equations in the static background.

The Einstein equations of $O(\varepsilon)$ are 
\begin{eqnarray}
&& \beta_1'' - \ddot{\beta}_1=0, \label{evol-1}\\ 
&&  \beta_0\psi_1'' +\beta_0'\psi_1'  - \beta_0\ddot{\psi}_1 +\beta_1\psi_0'' 
+ \beta_1'\psi_0'=0,  \\
&& 
\gamma_1'' -\ddot{\gamma}_1 +2\psi_0^{'}\psi_1'=0, \label{evol-3}\\
&& 
\beta_0'\beta_1''-2\beta_0^{'}\gamma_0'\beta_1'+\beta_0\psi_0^{'2} \beta_1'
+\beta_0'' \beta_1' + \beta_0'\psi_0^{'2} \beta_1
-\beta_0^{'2}\gamma_1' 
+2\beta_0\beta_0'\psi_0^{'}\psi_1' =0, \label{eq:1-order-constraint1}\\
&& \beta_0'\dot{\beta}_1' -  \beta_0 \psi_0^{'2} \dot{\beta}_1 
-\beta_0''\dot{\beta}_1 -\beta_0^{'2}\dot{\gamma}_1 
+2\beta_0\beta_0'\psi_0'\dot{\psi}_1   = 0. \label{eq:1-order-constraint2}
\end{eqnarray}
Eqs.~(\ref{evol-1})--(\ref{evol-3}) are the evolution equations for $\beta_1$, $\psi_1$ 
and $\gamma_1$, whereas Eqs.~(\ref{eq:1-order-constraint1}) 
and (\ref{eq:1-order-constraint2}) are the constraint equations which include 
no second order time derivative. 

In order to study the stability of the static AT-shell, 
we assume that there is no energy input into the AT-shell. Therefore 
we impose the outgoing wave boundary condition 
in which there is no ingoing gravitational 
wave from past null infinity. In the interior region, we impose the 
regularity condition at the symmetric axis. 
Firstly, we solve these equations in the interior and exterior regions by 
imposing the regularity condition at the symmetry axis and the outgoing 
wave boundary condition, and then
impose the junction conditions to glue these solutions at $\Sigma$. 

\subsection{Solutions for the interior region}

By using the expression of the static solutions 
in the interior region (\ref{eq:zero-interior-solution}), 
we rewrite the linearized Einstein equations for the interior region as
\begin{eqnarray}
&& \beta_1^{-''}-\ddot{\beta}_1^- = 0 \label{eq:beta-1-inner}\\
&& \psi_1^{-''} + \frac{1}{r} \psi_1^{-'} -\ddot{\psi}_1^- =0, \\
&& \gamma_1^{-''}-\ddot{\gamma}_1^- = 0,  \\
&& \beta_1^{-''} = \gamma_1^{-'},  \\
&& \dot{\beta}_1^{-'}=\dot{\gamma}_1^-. 
\end{eqnarray}
The last two equations can be integrated to give the relation
\begin{eqnarray}
\beta_1^{-'} +C^-= \gamma_1^- , \label{gamma1_in}
\end{eqnarray}
where $C^-$ is an integration constant. Therefore, $\gamma_1^-$ is easily obtained 
once $\beta_1^-$ is known. The others are three homogeneous equations. 
Since the background solution is static, we can assume that all  perturbations have a 
time dependence $e^{-i\omega_- t_-}$ and solve for $\omega_-$ in order to 
know whether complex or pure imaginary frequencies exist. 
The solutions for the three homogeneous equations are given by 
\begin{eqnarray}
&& \beta_1^-  = \mbox{Re}\left[ \ e^{-i\omega_- t_-}\left(
A^{\beta^-}(\omega_-) e^{i\omega_- r} + B^{\beta^-}(\omega_-) e^{-i\omega_- r}
\right)\right], \\
&& \psi_1^- = \mbox{Re}\left[  e^{-i\omega_- t_-} \left(
A^{\psi^-}(\omega_-) H_0^{(1)}(\omega_- r) + B^{\psi^-}(\omega_-) H_0^{(2)}(\omega_- r)
\right)\right], \\
&& \gamma_1^- = \mbox{Re}\left[ \ i\omega_- e^{-i\omega_- t_-} \left(
A^{\beta^-}(\omega_-) e^{i\omega_- r} - B^{\beta^-}(\omega_-) e^{-i\omega_- r}
\right) + C^-\right], 
\end{eqnarray}
where $A^{\beta^{-}}, B^{\beta^-}, A^{\psi^-}$ and $B^{\beta^-}$ are 
functions of $\omega_-$. \lq\lq Re\rq\rq denotes the real part, and $H^{(1)}_0$, 
$H^{(2)}_0$ are the Hankel functions of the first and second kind of order $0$, 
respectively. 

Now, we impose the regularity condition at the symmetric axis, $r=0$, in 
accordance with Hayward\cite{Hayward2000}. 
The regularity condition can be expressed in a geometrically invariant manner 
by using the norm of the Killing vectors, 
\begin{eqnarray}
\rho:=|\partial_{\phi} | = \beta e^{-\psi}, \quad \ell :=|\partial_z| = e^{\psi}.
\end{eqnarray}
At the symmetric axis, $\rho$ should vanish.
The axis is said to be regular if $\ell$ is finite and 
\begin{eqnarray}
&& \nabla \rho \cdot \nabla\rho = 1+\mathcal{O}(\rho^2), 
\label{eq:regular-cond-1}\\
&& \nabla\rho \cdot \nabla\ell = \mathcal{O}(\rho), 
\label{eq:regular-cond-2}\\
&& \nabla\ell \cdot \nabla\ell = \mathcal{O}(1), 
\end{eqnarray}
where $\nabla$ is the covariant derivative of the spacetime metric.
These regularity conditions are satisfied if and only if
\begin{eqnarray}
A^{\beta^-}+B^{\beta^-}  = 0, \quad  A^{\psi^-} -B^{\psi^-} =0, \quad 
 C^- =0.
\end{eqnarray}
Therefore, we obtain the regular interior solutions as
\begin{eqnarray}
&& \beta_1^-  = \mbox{Re}\left[e^{-i\omega_- t_-}
A^{\beta^-}(\omega_-) \left( e^{i\omega_- r} -  e^{-i\omega_- r}
\right)\right], \\
&& \psi_1^- = \mbox{Re}\left[2e^{-i\omega_- t_-} 
A^{\psi^-}(\omega_-) J_0(\omega_- r)\right], \\
&& \gamma_1^- = \mbox{Re}\left[i\omega_- e^{-i\omega_- t_-}
A^{\beta^-}(\omega_-) \left( e^{i\omega_- r} +  e^{-i\omega_- r}
\right)\right], 
\end{eqnarray}
where $J_0$ is the Bessel function of order $0$.
If $\omega_-$ is real and negative, we have to replace $J_0(\omega_- r)$ by 
$J_0(|\omega_-|r)$. However, since we are interested 
in complex or pure imaginary solutions for $\omega_-$, 
we need not treat such a case.

\subsection{Solutions for the exterior region}

Using the fact that $\psi_0^{'2}=-\gamma_0''$, 
the constraint equations (\ref{eq:1-order-constraint1}) 
and (\ref{eq:1-order-constraint2}) become 
\begin{eqnarray}
&& \beta_{1}^{+''}-(\gamma_{0}^{+'}\beta_{1}^+)'
+2\beta_0^+\psi_{0}^{+'}\psi_{1}^{+'}-\beta_{0}^{+'}\gamma_{1}^{+'}  =0,   \\
&&\dot{\beta}_{1}^{+'}- \gamma_{0}^{+'}\dot{\beta}_{1}^+ 
+2\beta_0^+ \psi_{0}^{+'} \dot{\psi}_1^+
- \beta_{0}^{+'}\dot{\gamma}_{1}^+ =0.
\end{eqnarray}
By noting that $\beta_0^+\psi_0^{+'}$ is constant, we can integrate
the above two equations and obtain 
\begin{eqnarray}
\gamma_1^+=  \frac{1}{\beta_0^{+'}}\left( \beta_1^{+'} -\gamma_0^{+'}\beta_1^+ 
+2\beta_0^+\psi_0^{+'} \psi_1^+ \right) + C^+, \label{gamma1_ex}
\end{eqnarray}
where $C^+$ is an integration constant. By using the background solutions 
(\ref{eq:0th-exterior-solution}), the evolution equations for 
$\beta_1^+, \psi_1^+$ and $\gamma_1^+$ become 
\begin{eqnarray}
&&\beta_1^{+''} -\ddot{\beta}_1^+=0, \\
 &&\psi_{1}^{+''} +\frac{1}{r}\psi_1^{+'}  -  \ddot{\psi}_1^+  
=  - \frac{1}{r}(\psi_0^{+'} \beta_1^+)', \label{psi-eq}\\
&& \gamma_1^{+''} -\ddot{\gamma}_1^+ +2\psi_0^{+'}\psi_1^{+'} =0.
\end{eqnarray}
Though the equation for $\psi_1^+$ is an inhomogeneous equation, 
one can easily see that $\psi_0^{+'}\beta_1^+$ is a particular solution to 
the equation\footnote{Note that $\psi_0^{+'} = -2y/r$ from 
Eq.~(\ref{eq:0th-exterior-solution})}.
By the same reason as the interior solutions, 
we can assume the time dependence $e^{-i\omega_+ t_+}$ for the exterior solutions. 
The outgoing wave solutions for the above equations are written as 
\begin{eqnarray}
&& \beta_1^+ =  \mbox{Re}\left[A^{\beta^+}(\omega_+)e^{-i \omega_+ (t_+-r)}\right],   \\
&& \psi_1^+ = g_{\psi}^+  -\frac{2y \beta_1^+}{r}, \\
&& \gamma_1^+ =  \beta_1^{+'}
 +\frac{4y^2}{r} \beta_1^+ -4y g_{\psi}^+ + C^+,
\end{eqnarray}
where we have introduced the outgoing homogeneous solution for Eq.~(\ref{psi-eq}) as
\begin{equation}
g_{\psi}^+ :=  \mbox{Re}\left[A^{\psi^+}(\omega_+)e^{-i\omega_+ t_+} 
H_0^{(1)}(\omega_+ r)\right]. 
\end{equation}
If $\omega_+$ takes a negative real value, we have to replace 
$H_0^{(1)}(\omega_+ r)$ by $H_0^{(2)}(|\omega_+|r)$ 
so as to keep the outgoing wave condition. 
However, since we are interested in complex or pure imaginary solutions 
for $\omega_+$, we need not treat such a case. 

\subsection{Solving the junction conditions at $\Sigma$}

Now, we glue the interior solution to the exterior one at $\Sigma$.
For that purpose, we have to relate the time coordinates $t_-$ and $t_+$ 
at $\Sigma$. 
%which are arguments of the interior and exterior solutions, respectively. 
Firstly, we relate $t_-$ and $t_+$ to $\tau$. 
From Eq.~ (\ref{eq:X}) and the comoving condition 
$V=0$, we have 
\begin{eqnarray}
\frac{d t_-}{d \tau} = e^{-\gamma^-+\psi^-} 
= e^{\psi_s}+\varepsilon e^{\psi_s} (\psi_1^--\gamma_1^-)+{\cal O}(\varepsilon^2).
\label{eq:dt^--dtau}
\end{eqnarray}
The second term in the right hand side in the above equation contributes 
in all relevant equations only of ${\cal O}(\varepsilon^2)$, since 
the background solution does not depend on $t_-$. Therefore, 
we may set, in the linearized equations, 
$t_- = e^{\psi_s} \tau$.
Similarly, for the exterior time coordinate, we may set
$t_+ = (1+2y)^{-2} e^{\psi_s}\tau$. 
%The relations (\ref{eq;t^--tau}) and (\ref{eq;t^+-tau}) 
%are justified in first order equations 
%only when the leading order of perturbation is static.
From these relations, we obtain 
\begin{eqnarray}
t_+ = (1+2y)^{-2} t_- ~~~{\rm on}~~\Sigma.  \label{t-relation}
\end{eqnarray}
%With this relation, the interior solutions can be matched to the 
%exterior solutions. 

The continuity of $\beta$ and $\psi$ at $\Sigma$ gives 
\begin{eqnarray}
&& \beta_1^-|_{r=R} = \beta_1^+|_{r=R}, \label{eq:continuity-beta-1} \\
&& \psi_1^-|_{r=R} = \psi_1^+|_{r=R}. \label{eq:continuity-psi-1}
\end{eqnarray}
The above conditions imply that the time dependences of solutions for 
$\beta_1$ and $\psi_1$ in $M_\pm$ are the same at $\Sigma$, and hence 
$\omega_+ t_+=\omega_- t_-$ should be satisfied.  
Then from Eq.~(\ref{t-relation}), we have $\omega_+=(1+2y)^2\omega_-$.
Hereafter, for notational simplicity, 
we abbreviate $\omega_-$ as $\omega$, or equivalently,  
\begin{eqnarray}
\omega_-&=&\omega, \\
\omega_+&=&(1+2y)^2\omega.
\end{eqnarray}

The junction conditions of $O(\varepsilon)$ are obtained by directly expanding 
Eqs.~(\ref{eq:junction-1}), (\ref{eq:junction-2}) and (\ref{eq:junction-3}) with respect to 
$\varepsilon$ as
\begin{eqnarray}
&&\left[e^{-\gamma_0}(\psi_1'- \gamma_1 \psi_0')  \right]^+_-
= \frac{2y}{R(1+2y)^2}\left(
\frac{1}{1+y}\frac{\beta_1^-}{R}+\frac{1+2y}{1+y}\psi_1^- \right)\biggr|_{r=R}, 
\label{eq:junction-first-psi} \\
&&\left[e^{-\gamma_0}(\gamma_1'- \gamma_1  \gamma_0')  \right]^+_-
= \frac{4y^2}{R(1+2y)^2}\left(\frac{1}{1+y} \psi_1^- 
-\frac{3+2y}{1+y}\frac{\beta_1^-}{R} \right)\biggr|_{r=R},  
\label{eq:junction-first-gamma} \\
&&\left[e^{-\gamma_0}\left(\beta_1' - \gamma_1 \beta_0' 
\right)  \right]^+_-
= \frac{4y(1+y)}{(1+2y)^2}\left(\frac{1}{1+y} \psi_1^- 
+ \frac{y}{1+y}\frac{\beta_1^-}{R} \right)\biggr|_{r=R}.
\label{eq:junction-first-beta}
\end{eqnarray}
%The right hand sides of the above conditions 
%are evaluated at the shell, where 
Since $\beta_1$ and $\psi_1$ are continuous at $r=R$, 
one may replace $\beta_1^-$ and $\psi_1^-$ in the right hand 
side of the above equations with $\beta_1^+$ and $\psi_1^+$, respectively.
By substituting Eq.~(\ref{gamma1_in}) with $C^-=0$ and Eq.~(\ref{gamma1_ex}) 
into Eq.~(\ref{eq:junction-first-beta}), and 
by using the static solutions (\ref{eq:zero-interior-solution}) and 
(\ref{eq:0th-exterior-solution}), we obtain 
%The equations (\ref{eq:continuity-beta-1}), (\ref{eq:continuity-psi-1}) and 
%(\ref{eq:junction-first-beta}) give
\begin{eqnarray} 
C^+ =0.
\end{eqnarray}
Then, four equations (\ref{eq:continuity-beta-1}), 
(\ref{eq:continuity-psi-1}), (\ref{eq:junction-first-psi}) and
 (\ref{eq:junction-first-gamma}) 
determine four amplitudes $A^{\beta^{\pm}}$ and $A^{\psi^{\pm}}$.

The continuity conditions (\ref{eq:continuity-beta-1}) and 
(\ref{eq:continuity-psi-1}) give
\begin{eqnarray}
&& (1+2y)^2e^{i\omega \hat{R}} \frac{A^{\beta^+}}{R}
- 2i\frac{A^{\beta^-}}{R} \sin \omega R  = 0,  \label{eq:continuity-beta-2}\\
&& (1+2y)^2 A^{\psi^+} H_0^{(1)}(\omega \hat{R}) -{2y} 
 (1+2y)^2e^{i\omega \hat{R}}\frac{A^{\beta^+}}{R}
-2A^{\psi^-} J_0(\omega R) =0, \label{eq:continuity-psi-2}
\end{eqnarray}
where, for notational simplicity,  
we have introduced a new quantity $\hat{R} = (1+2y)^2 R$. 
 The junction conditions (\ref{eq:junction-first-psi}) and 
(\ref{eq:junction-first-gamma}) lead
\begin{eqnarray}
&& -\omega R(1+2y)^4 A^{\psi^+} H_1^{(1)}(\omega \hat{R}) 
-{8y^2}(1+2y)^2A^{\psi^+}H_0^{(1)}(\omega \hat{R})
+2(1+2y)^2 \omega  RA^{\psi^-}J_1(\omega R)  \nonumber \\
&& +{2y}(1+4y^2) 
(1+2y)^2e^{i\omega \hat{R}} \frac{A^{\beta^+}  }{R}
- \frac{4iy}{(1+y)} \frac{A^{\beta^-}}{R}  \sin \omega R  
 -\frac{4y(1+2y)}{(1+y)} A^{\psi^-}J_0(\omega R)  =0, \nonumber \\
\label{eq:junction-first-psi-2}
\end{eqnarray}
and 
\begin{eqnarray}
&& -\omega^2 R^2(1+2y)^6 e^{i\omega \hat{R}} \frac{A^{\beta^+}}{R} 
-{4y^2(1+4y^2)}(1+2y)^2e^{i\omega \hat{R}}  \frac{A^{\beta^+} }{R}  
+4y(1+2y)^4\omega R A^{\psi^+}H_1^{(1)}(\omega \hat{R}) 
\nonumber \\
&& 
+{16y^3}(1+2y)^2A^{\psi^+}H_0^{(1)}(\omega \hat{R}) + 2i(1+2y)^2 \omega^2 R^2
\frac{A^{\beta^-}}{R} \sin \omega R  
-\frac{8y^2}{(1+y)} A^{\psi^-}J_0(\omega R)  \nonumber \\
&& + \frac{8iy^2(3+2y)}{(1+y)}
\frac{A^{\beta^-}}{R} \sin \omega R  =0,
\label{eq:junction-first-gamma-2}
\end{eqnarray}
respectively.
These equations can be put into the following matrix form:
\begin{eqnarray}
\sum_{j=1}^{4}M_{ij} A^j =0, \label{A-eq}
\end{eqnarray}
where $i=1,2,3,4$ and we have defined
\begin{eqnarray}
A^1 := A^{\psi^+}, \quad A^2 := \frac{A^{\beta^+}}{R}, \quad A^3 := A^{\psi^-},
\quad A^4 := \frac{A^{\beta^-}}{R}.
\end{eqnarray}
The components of the matrix $M_{ij}$ are given by 
\begin{eqnarray}
&& M_{11} = M_{13} = M_{24} =0, \\
&& M_{12} =(1+2y)^2e^{i\omega \hat{R}}, \\
% && M_{13} = 0\\
&& M_{14} = -2i\sin \omega R, \\
&& M_{21} = (1+2y)^2 H_0^{(1)}(\omega \hat{R}), \\
&& M_{22} = -{2y}  (1+2y)^2e^{i\omega \hat{R}}, \\
&& M_{23} =-2J_0(\omega R),  \\
% && M_{24} = 0, \\
&& M_{31} = -\omega R(1+2y)^4 H_1^{(1)}(\omega \hat{R}) 
-{8y^2}(1+2y)^2 H_0^{(1)}(\omega \hat{R}), \\
&& M_{32} =  2y(1+4y^2) (1+2y)^2e^{i\omega \hat{R}}, \\
&& M_{33} = 2(1+2y)^2 \omega R J_1(\omega R)- \frac{4y(1+2y)}{(1+y)} J_0(\omega R), \\
&& M_{34} = - \frac{4i y}{(1+y)} \sin \omega R, \\
&& M_{41} = 4y(1+2y)^4\omega R H_1^{(1)}(\omega \hat{R}) 
+{16y^3}(1+2y)^2H_0^{(1)}(\omega \hat{R}) , \\
&& M_{42} = -\omega^2R^2 (1+2y)^6 e^{i\omega \hat{R}}
- {4y^2(1+4y^2)}(1+2y)^2e^{i\omega \hat{R}}, \\
&& M_{43} = -\frac{8y^2}{(1+y)} J_0(\omega R) , \\
&& M_{44} = 2i (1+2y)^2 \omega^2 R^2 \sin \omega R +\frac{8iy^2(3+2y)}{(1+y)} 
\sin \omega R.
\end{eqnarray}
%The condition that nonzero solutions exist is that the determinant  
In order that there are non-trivial solutions for Eq.~(\ref{A-eq}), 
the following equation should be satisfied; 
\begin{eqnarray}
\mbox{det}\left(M_{ij} \right) =0.
\label{eq:matrix-det}
\end{eqnarray}
For fixed $y$ and $R$, this is an equation for frequency $\omega$.
Here, instead of (\ref{eq:matrix-det}), we consider a real equation
\begin{eqnarray}
|\mbox{det}\left(M_{ij} \right)|^2 =0.
\label{eq:condition-omega}
\end{eqnarray}
Using a dimensionless quantity $z=\omega R$, Eq.~(\ref{eq:condition-omega}) 
is rewritten in the form 
\begin{eqnarray}
\frac{256y^2(1+2y)^{12}}{(1+y)^2}
F(z) \sin^2z=0,
\label{eq:determinant}
\end{eqnarray}
where $F(z)$ is defined by 
\begin{eqnarray}
F(z) &:=& \Bigg[ \bigg(2y(1+y)\left((1+6y+4y^2)z^2-2y  \right) J_0(z) \nonumber \\ 
 && \qquad +z\left(y(2+y(1+2y)^2) -(1+y)^2(1+2y)^2z^2 \right) J_1(z) 
\bigg) H_0^{(1)}\big((1+2y)^2z\big) \nonumber \\
&&\qquad  +(1+2y)^2z 
\bigg( (z^2+y(2+y)(z^2-1))J_0(z)+2y(1+y)z J_1(z)\bigg)H_1^{(1)}\big((1+2y)^2z\big)\Bigg]
\nonumber \\
&& \Bigg[
\bigg(2y(1+y)\left((1+6y+4y^2)z^2-2y  \right) J_0(z)  \nonumber \\
&&\qquad +z\left(y(2+y(1+2y)^2) -(1+y)^2(1+2y)^2z^2 \right) J_1(z) 
\bigg)H_0^{(2)}\big((1+2y)^2z\big) \nonumber \\
&& +(1+2y)^2z \bigg( (z^2+y(2+y)(z^2-1)J_0(z) +2y(1+y)zJ_1(z)  ) \bigg) 
H_1^{(2)}\big((1+2y)^2z\big)
\Bigg]. \nonumber \\
\end{eqnarray}

It is easily seen  that, for integer $n$, there are infinitely large numbers of  
real solutions $z=n\pi$, or equivalently, 
\begin{eqnarray}
\omega  = \frac{n\pi}{R}.
\end{eqnarray}
These are radial oscillation modes of the shell around the static configuration.
The existence of these oscillation modes is compatible with the C-energy 
argument given by Apostolatos and Thorne\cite{AT}.
The gravitational emission may cause the damping of the oscillation, but this effect 
appears in ${\cal O}(\varepsilon^2)$.

\subsection{Unstable modes}

In this subsection, we look for zero points of the function $F(z)$
in complex $z$ plane. We solve the equation 
\begin{eqnarray}
F(z)=0,
\label{eq:F=0}
\end{eqnarray}
by using \lq\lq FindRoot\rq\rq program in Mathematica (version 7.0). 
% and obtain complex solutions. 
As mentioned, the parameter $y$ characterizes the static solution, 
and hence we search for the solutions of $\omega$ as a function of $y$. 
%investigate the cases of several values of $y$.

Since we invoked the numerical method to find the solutions, 
we have investigated a limited 
number of points in the domain $0\leq y <\infty$, not all $y$. 
But, as far as we have investigated, there exists at least one unstable 
mode for each $y$. Therefore, it is reasonable to conclude that 
there are unstable modes for all $y\geq0$. 
The angular frequency $\omega$ of the unstable solution is written in the form
\begin{equation}
\omega=\omega_{\rm R}(y)+i\omega_{\rm I}(y),
\end{equation} 
where $\omega_{\rm R}$ is a real function of $y$, and $\omega_{\rm I}$ is 
also a real function of $y$ but is positive.   
We also found that if $\omega=\omega_{\rm R}+i\omega_{\rm I}$ is a solution, then 
$\omega=\omega_{\rm R}-i\omega_{\rm I}$ is also a solution. 

\begin{figure}[htbp]
  \centering
  \includegraphics[width=0.65\linewidth,clip]{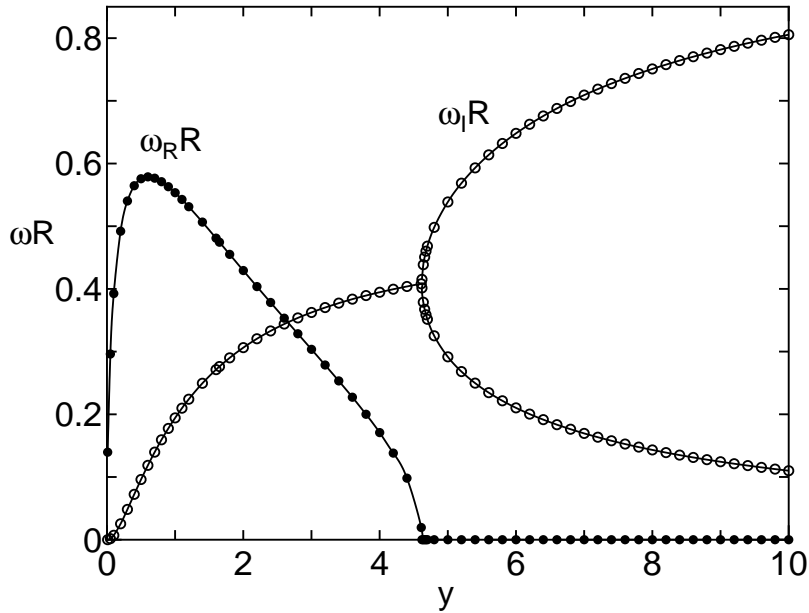}
  \caption{The solutions of the first class are depicted.
For $y\gtrsim  4.61$, the complex modes in this class become pure imaginary.}
  \label{fig:1st-class}
\end{figure}

\begin{figure}[htbp]
  \centering
  \includegraphics[width=0.65\linewidth,clip]{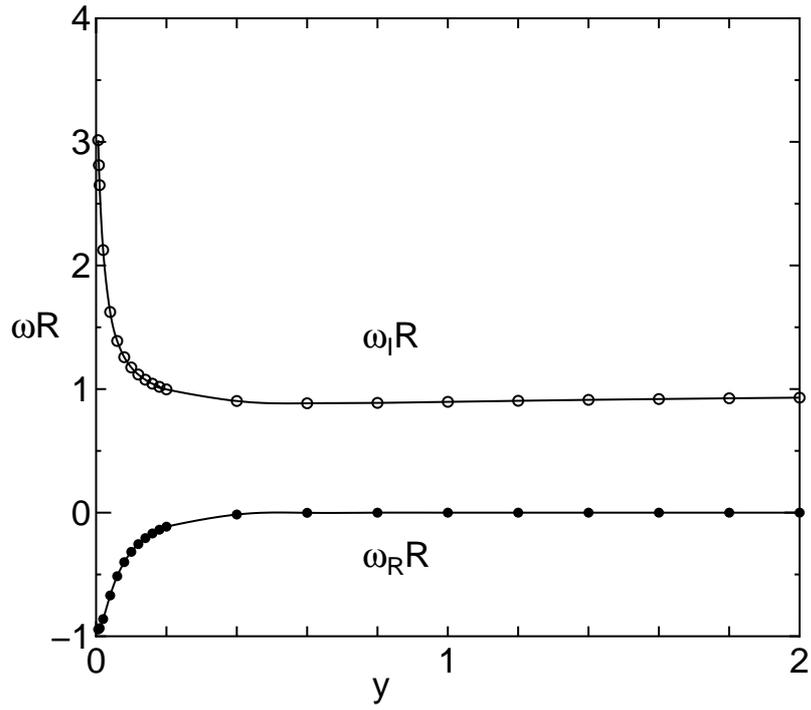}
  \caption{The solutions of the second class are depicted. 
For $y\gtrsim 1.4$, the complex modes in this class become pure imaginary.}
  \label{fig:2nd-class}
\end{figure}

\begin{figure}[htbp]
  \centering
  \includegraphics[width=0.6\linewidth,clip]{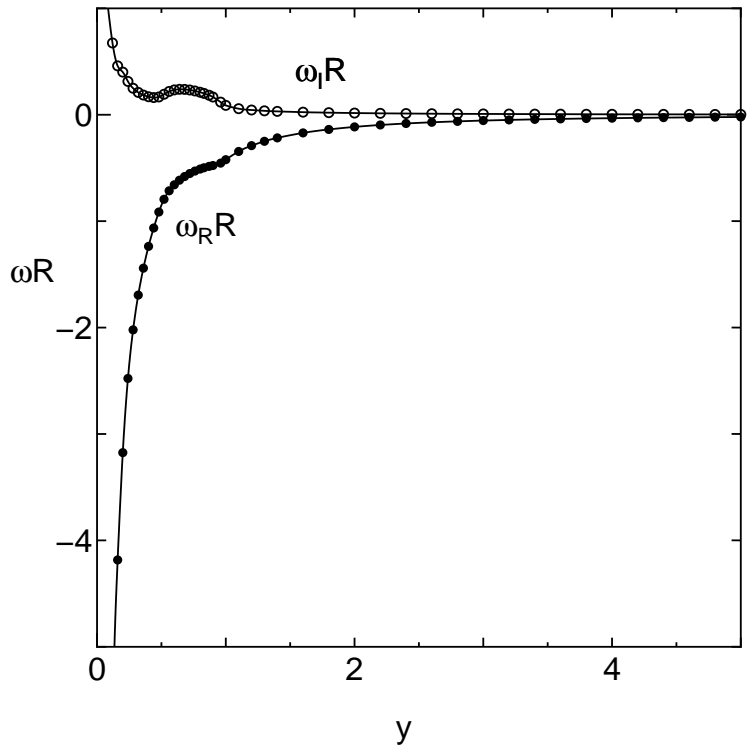}
  \caption{The solutions of the sixth class are depicted.  }
  \label{fig:6th-class}
\end{figure}

We have found sixth classes of the unstable solutions. 
We depict $\omega_{\rm R}$ and $\omega_{\rm I}$ of the first class as functions 
of $y$ in Fig.~\ref{fig:1st-class}, 
whereas those of the second, third, fourth, fifth and sixth 
% we depict the same as well but of the second, third, fourth and fifth 
classes are dipicted in 
Figs.~{\ref{fig:2nd-class}}--{\ref{fig:6th-class}}, respectively.
%Fig.~{\ref{fig:2nd-class}}, Fig.~\ref{fig:3rd-class}, 
%Fig.~\ref{fig:4th-class}, Fig.~\ref{fig:5th-class} 
%and Fig.~{\ref{fig:6th-class}}, respectively.
%%%%%%%%%%%%%%%%%%%%%%%%%%
%%%%%%%%%%%%%%%%%%%%%%%%%%

%%%%%%%%%%%%%%%%%%%%%%%%%%
%%%%%%%%%%%%%%%%%%%%%%%%%%

In the case of the first class, the real part $\omega_{\rm R}$ increases from zero 
for $0\leq y\lesssim0.6$, decreases for $0.6\lesssim y \lesssim4.61$ and then becomes 
very tiny value $\epsilon$ for $y\gtrsim4.61$. The maximum value of $\omega_{\rm R}$ 
is about $0.579R^{-1}$ at $y=0.6$. The tiny value $\epsilon$ 
depends on initial values in numerical investigation, and 
therefore $\epsilon$ will be a numerical error.
Hence, it is reasonable to conclude that the numerical solutions with the form of
$\omega =\epsilon \pm i\omega_I $ correspond to pure imaginary solutions.
The imaginary part $\omega_{\rm I}$ 
of the first class increases from zero for $0\leq y\lesssim4.61$ and bifurcates 
at $y \simeq 4.61$: one sequence monotonically increases and the other 
monotonically decreases for $4.61 \lesssim y$.
% The maximum value of $\omega_{\rm I}$ is about $0.408R^{-1}$. 
At $y\simeq 4.61$, the derivatives of both $\omega_{\rm R}$ and 
$\omega_{\rm I}$ with respect to $y$ are discontinuous.

In the case of the second class, the real part
$\omega_{\rm R}$  increases for $y \lesssim 1.4$ and becomes 
very tiny value for $y\gtrsim 1.4$. 
By the same reason as in the first class, 
it is reasonable to conclude that these solutions are pure imaginary. 
The imaginary part 
$\omega_{\rm I}$ decreases for $y \lesssim 0.8$ 
and increases from 0.89$R^{-1}$ for $0.8\lesssim y $, 
approaching $R^{-1}$ as $y\rightarrow \infty$. 
The imaginary part $\omega_{\rm I}$ of the second class is larger 
than that of the first class.

%%%%%%%%%%%%%%%%%%%%%%%%%%
\begin{figure}[htbp]
  \centering
  \includegraphics[width=0.65\linewidth,clip]{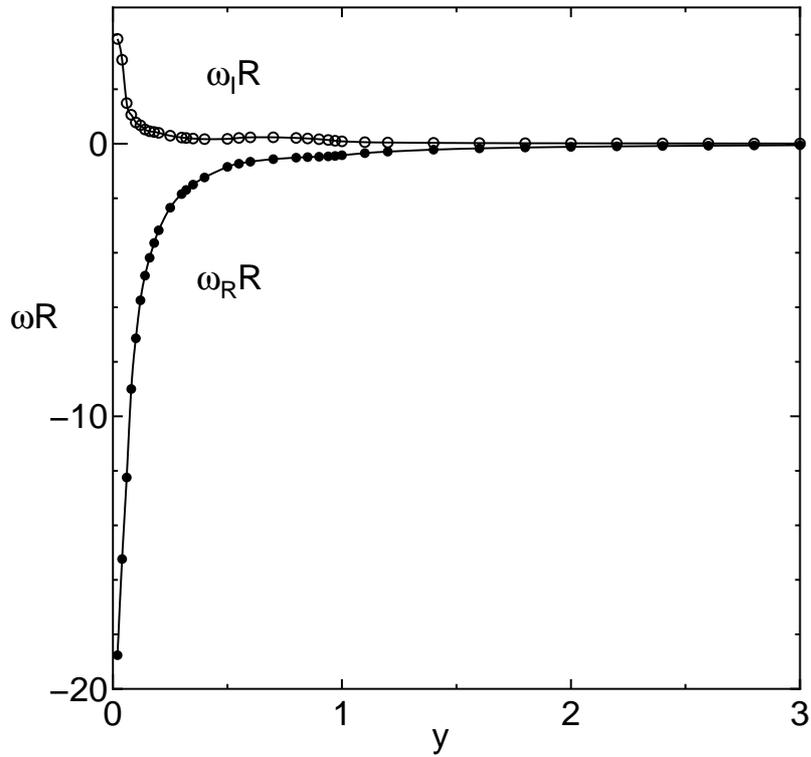}
  \caption{The solutions of the third class are depicted.}
  \label{fig:3rd-class}
\end{figure}
%%%%%%%%%%%%%%%%%%%%%%%%%%
In the case of the third class, the real part $\omega_{\rm R}$ monotonically increases 
and approaches 0 as $y$ increases.
The imaginary part 
$\omega_{\rm I}$ decreases for $y \lesssim 0.4$
and increases from 0.169$R^{-1}$ to 0.236$R^{-1}$ 
for $0.4 \lesssim y \lesssim 0.7$.
For $0.7\lesssim y$, $\omega_{\rm I}$ decreases and 
approaches 0 as $y$ increases.

%%%%%%%%%%%%%%%%%%%%%%%%%%
\begin{figure}[htbp]
  \centering
  \includegraphics[width=0.65\linewidth,clip]{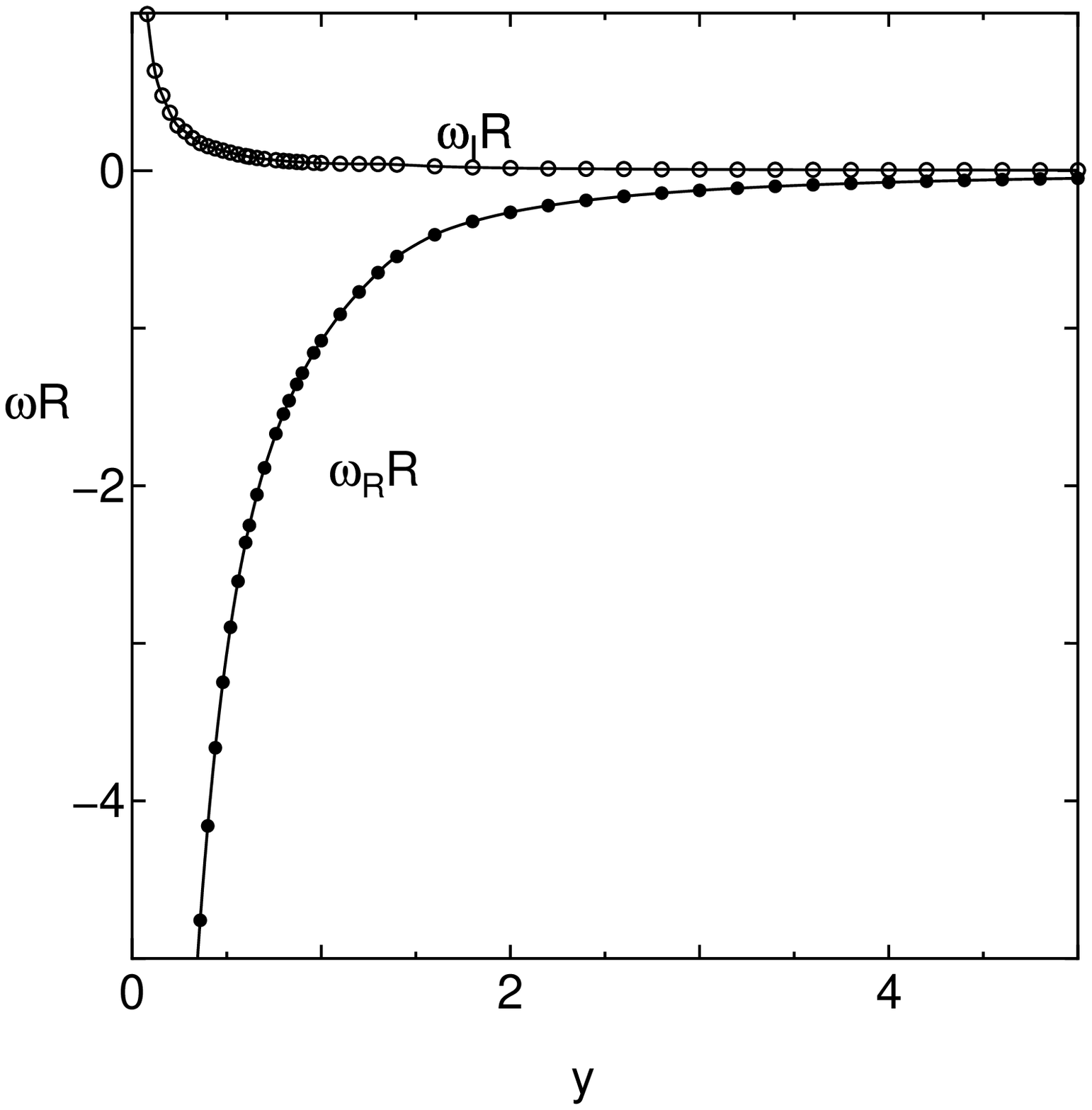}
  \caption{The solutions of the fourth class are depicted.  }
  \label{fig:4th-class}
\end{figure}
%%%%%%%%%%%%%%%%%%%%%%%%%%

%%%%%%%%%%%%%%%%%%%%%%%%%%
\begin{figure}[htbp]
  \centering
  \includegraphics[width=0.63\linewidth,clip]{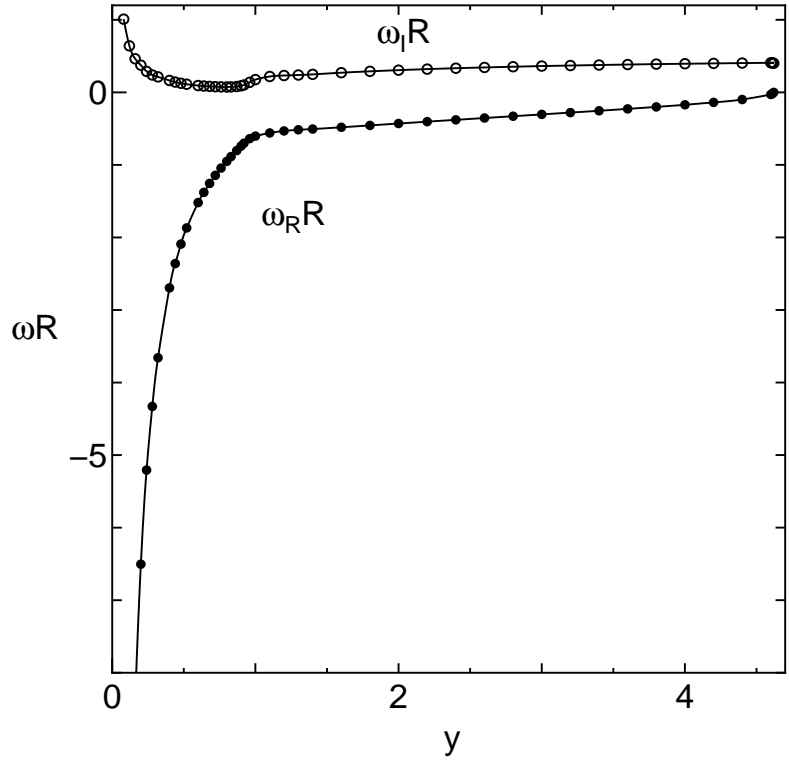}
  \caption{The solutions of the fifth class are depicted. 
At $y\simeq 4.62$, the solution in this class becomes pure imaginary
and this class seems to connect to the pure imaginary modes 
in the first class.}
  \label{fig:5th-class}
\end{figure}
%%%%%%%%%%%%%%%%%%%%%%%%%%

%%%%%%%%%%%%%%%%%%%%%%%%%%
%%%%%%%%%%%%%%%%%%%%%%%%%%

In the case of fourth class, the real part $\omega_{\rm R}$ monotonically increases. 
The imaginary part $\omega_{\rm I}$ monotonically decreases
and approaches 0.

The fifth class has monotonically increasing $\omega_{\rm R}$ for $y \lesssim 4.61$. 
At $y\simeq 4.62$, $\omega_{\rm R}$ becomes very tiny value, 
and for $y \gtrsim 4.62$ the solution becomes pure imaginary.
The imaginary part $\omega_{\rm I}$ 
monotonically decreases to 0.075$R^{-1}$ for $y \lesssim 0.8$ and monotonically
increases for $0.8 \lesssim y \lesssim 4.61$.
At $y=4.62$, the solution coincides with that of the first class. 
Therefore, this class seems to connect to the pure imaginary modes in the first class for 
$y\gtrsim 4.62$. 

In the case of sixth class, the real part $\omega_{\rm R}$ monotonically increases
and approaches 0.
The imaginary part $\omega_{\rm I}$ 
monotonically decreases to 0.016$R^{-1}$ for $y \lesssim 0.44$ 
and increases to 0.24$R^{-1}$ for $0.44 \lesssim y \lesssim 0.64$.
For $0.64\lesssim y$, $\omega_{\rm I}$ decreases and approaches 0.

In order to understand what the existence of the unstable modes implies, 
we investigate the circumferential radius ${\cal R}$ of the shell. 
The perturbation of $O(\varepsilon)$ for the circumference radius of the AT shell is 
\begin{eqnarray}
{\cal R}_1
&=&(\beta_1 e^{-\psi_0}-\beta_0\psi_1)|_{r=R} \nonumber \\
&=&\mbox{Re} \biggl[e^{-i\omega t_-}
\Bigl[  (e^{-\psi_s}+y)e^{i\omega \hat{R}}A^{\beta^+}(\omega) 
-RH_0^{(1)}(\omega \hat{R})A^{\psi^+}(\omega)
 \Bigr ] \biggr].
\end{eqnarray}
From the above equation, we can see that 
${\cal R}_1$ is written symbolically  in the form 
\begin{eqnarray}
{\cal R}_1=\mbox{Re} \left[
e^{-i \omega t_-}\left(C_1+iC_2\right)
\right]
&=& \sqrt{C_1^2+C_2^2} ~e^{\omega_I t_-} \cos(\omega_R t_- - \phi),
\label{eq:behavior-of-radius}
\end{eqnarray}
where $C_1$ and $C_2$ are real numbers, and 
we have introduced a constant $\phi:=\tan^{-1}C_2/C_1$.
The above equation implies that, if $\omega_{\rm R}$ does not vanish, 
the circumference radius oscillates and its amplitude grows exponentially with time. 
If $\omega_{\rm R}$ vanishes, there is a mode of which the shell does 
not oscillate radially but just expands or contracts exponentially. 
In both cases, the static shell is unstable up to the linear order. 

When $\omega_R=0$, whether the AT-shell expands or contracts is 
determined by initial condition.
% In order to know whether the AT-shell expands or contracts in case of $\omega_R=0$, 
% we investigate sign of ${\cal R}_1$: 
% if ${\cal R}_1$ is negative, the AT-shell will contract exponentially, 
% and by contrast, if ${\cal R}_1$ is positive, the AT-shell will expand exponentially.
% Hence, the sign of ${\cal R}_1$ is what we want to know.
In this case, ${\cal R}_1$ can be expressed as 
\begin{eqnarray}
{\cal R}_1 = e^{\omega_I t_-} C_1.
\end{eqnarray}
Therefore, if ${\cal R}_1$ is initially positive, then $C_1$ is positive 
and we find that the AT-shell expands exponentially.
By contrast, if ${\cal R}_1$ is initially negative, then $C_1$ is negative
and it contracts exponentially. 
This behavior of the AT-shell can be understood as infinite period 
limit ($T=2\pi/\omega_R \to \infty$) of the oscillation behavior 
(\ref{eq:behavior-of-radius}).

The appearance of the pure imaginary modes depends on the parameter $y$.
By using the speed of orbital motion of each constituent particle 
measured by an observer who rides on the static AT shell, say $v$, 
the parameter $y$ can be expressed as\cite{AT} 
\begin{eqnarray}
y= \frac{v^2}{1-v^2}. 
\end{eqnarray}
It is easy to see that $y$ is a monotonically increasing function of $v$.
If the velocity $v$ of each constituent particle is smaller than a critical value 
$v_{\rm c} \simeq 0.76$ which corresponds to $y\simeq1.4$, 
the amplitude of the radial oscillation grows, whereas if 
$v>v_{\rm c}$,  circumferential radius will expands or contracts 
exponentially with time. 

%==============================================
\section{summary and discussion}
\label{section-summary}
%===========================================

We have investigated linear perturbation of the AT-shell around the static solution,
and have found that the static state is unstable in the sense that,
if the speed of orbital motion of the constituent particles (denoted by $v$) is small, 
the perturbation of the shell's circumference radius 
oscillates with exponentially growing amplitude
and the shell does not settle down into the static configuration,
and if $v$ is larger than the critical value $v_{\rm c}\simeq 0.76$, 
the shell just expands or contracts exponentially 
with time. Whether the shell expands or contracts for large velocity 
depends on the initial condition. If initially the circumference radius 
is larger than that of the static shell, then the AT-shell just expands exponentially. 
By contrast, if initially it is smaller than that of the static shell, 
the AT-shell just contracts exponentially.

Since we invoked the numerical method to find the solutions, 
we have investigated a limited 
number of points in the domain $0\leq y <\infty$, not all $y$. 
But, as far as we have investigated, there exists at least one unstable 
mode for each $y$. Therefore, it is reasonable to conclude that the static 
AT-shell is unstable for all $y$ up to $O(\varepsilon)$.

This result is compatible with the previous work given in Ref.~[\ref{NIK}] 
which showed that there exist momentarily static 
and radiation free initial states of the AT-shell which do not 
settle down into the static state, unless gravitational radiation extracts 
an infinite amount of C-energy from the AT-shell. 
More concretely, in this study, it was shown  that,
if the initial circumference radius of the AT-shell 
is greater than that of the expected final static state (having
fixed $\alpha, \lambda, R$ and $\psi_s$), 
then the AT-shell cannot settle down into the static state,
and if the initial circumference radius of the AT-shell is smaller than 
that of the static state,
then it is not forbidden that the AT-shell settles down into the equilibrium
static configuration. 
However, the existence of the unstable modes in our perturbation analysis
does not depend on whether 
the initial circumference radius is greater than that of the static state or not.
It means that, 
even if the initial circumference is smaller than that of 
the expected static shell, 
the AT-shell does not settle down into the static state.

Hamity, C$\acute{\mbox{e}}$cere and  Barraco\cite{Hamity:2007jh} 
discussed the stability of the AT-shell
and concluded that for $y<0.78049$ ($\sqrt{y}<0.8836$), the static 
configuration is stable and for $y>0.78049$ ($\sqrt{y}>0.8836$),
it is unstable. Their result seems to be inconsistent to Ref.[\ref{NIK}], 
and further, in the present analysis, we do not find any evidence for it. 
They investigated only the sequence of the static configuration 
%characterized by $\Lambda_{eq}(y)$ 
by a Gedanken experiment.  
However, it should be noted that in order to obtain a definite conclusion 
on the stability of the static AT-shell, 
it would be necessary to solve the Einstein equations. 
In contrast to Hamity et al., we have solved the Einstein equations, 
although we have used linear approximation. 

Our perturbation analysis shows that the static AT-shell solution 
is unstable but does not say anything about the shell's finial state.
To make it clear, perturbation theory will not be adequate and 
we will have to solve full equations by using numerical relativity.
These will be future works.

%%%%%%%%%%%%%%%%%%%%%%%%%%%%%%%%%%%%%%%%%%%%%%%%%%%%%%%%%%%%%%%%%
\section*{Acknowledgments}
It is our pleasure to thank Hideki Ishihara for his valuable discussion. 
We are also grateful to colleagues in the astrophysics and gravity 
group of Osaka City University for helpful discussion and criticism.
% \end{acknowledgments}
%%%%%%%%%%%%%%%%%%%%%%%%%%%%%%%%%%%%%%%%%%%%%%%%%%%%%%%%%%%%%%%%%

\end{document}